\documentclass[twocolumn,showpacs,preprintnumbers,amsmath,amssymb]{revtex4}

\usepackage{graphicx}

\providecommand*{\bra}[1]{\langle#1\rvert}
\providecommand*{\ket}[1]{\lvert#1\rangle}

\begin{document}

\title{Explicit tensor network representation for the ground states of
       string-net models}

\author{Oliver Buerschaper}
\author{Miguel Aguado}
\affiliation{Max-Planck-Institut f\"ur Quantenoptik\\Hans-Kopfermann-%
Stra\ss e 1, 85748 Garching, Germany}
\author{Guifr\'e Vidal}
\affiliation{School of Physical Sciences, The University of Queensland\\
Brisbane, QLD, 4072, Australia}

\date{\today}

\begin{abstract}
    The structure of string-net lattice models, relevant as examples of
    topological phases, leads to a remarkably simple way of expressing their
    ground states as a tensor network constructed from the basic data of the
    underlying tensor categories. The construction highlights the importance of
    the fat lattice to understand these models.
\end{abstract}

\pacs{}

\maketitle

\section{\label{sec:intro}Introduction}

Topological phases of lattice models recently attracted interest because of
their potential use as quantum memories and quantum computers \cite{Kitaev}.
While it is unclear whether their shortcomings at finite temperature can be
satisfactorily resolved \cite{finiteT}, these models remain appealing as systems with topological order
\cite{WenOrder} and quantum error correcting codes. The first such proposal was
Kitaev's paper \cite{Kitaev}, where the Abelian toric code was introduced, and a
class of non-Abelian generalisations, the quantum double models, with roots in
the theory of Hopf algebras. These are examples of gauge models with discrete
gauge group \cite{Wegner, Bais}. More recently, Levin and Wen \cite{LevinWen}
introduced their string-net models, arguing for string-net condensation as a
basic mechanism underlying topological order. A crucial device in the
formulation of these models is the fat lattice, which allows an interpretation
of the lattice model in terms of a theory in the continuum, and provides insight
in the role of the operators constituting the Hamiltonian.

In order to have a handle on the properties of topological phases on the
lattice, it is desirable to have as much theoretical control as possible over the form of
topological states. Accordingly, explicit tensor network representations for the ground state of string-net models have been recently presented. Thus, Ref.~\cite{VerstraetePower} describes a \emph{projected entangled-pair state} (PEPS) representation of the toric code and resonating valence bond states, whereas a \emph{multi-scale entanglement renormalization ansatz} (MERA) representation of all quantum doubles \cite{AguadoVidal} and all string-net models \cite{Koenig} is also known. Finally, Ref.~ \cite{GuTopo} discusses a \emph{double line} tensor network representation of the $\mathbb{Z}_2$ gauge model and the double-semion model.

In this work we describe a simple construction expressing the ground states of
any string-net model as a tensor network constructed with $F$-tensors. The starting
point is the form of the Hamiltonian as a sum of commuting projectors,
$H=-\sum_i P_i$, which leads to the realisation of the ground level as the $+1$
eigenspace of the product $\prod_i P_i$. The tensor network follows in a
remarkably straightforward way. Its form is reminiscent of a classical
statistical mechanical partition function with local (albeit possibly complex) weights, which is why we
call it a Boltzmann weight tensor network. The needed ingredients are the
data of the underlying tensor category as explained in \cite{LevinWen}, i.\,e.,
the fusion rules and associated $F$-tensors. The construction is most
appropriately understood from the fat lattice perspective.

In section \ref{sec:stringnets}, we provide a short exposition to string-net
models and the fat lattice picture. In section \ref{sec:networks},
the Boltzmann weight tensor network construction for string-net
ground states is presented step by step. Essentially,
we use the projectors on the ground level in order to
build a simple expression for the ground state using $F$-tensors.

The Boltzmann weight tensor network differs from the MERA of Ref.~\cite{Koenig}, also written in terms of $F$-tensors, in that it is much simpler than the latter---e.\,g., it is a two-dimensional network, while the MERA spans three dimensions. On the other hand, although conceived independently, our construction coincides for the $\mathbb{Z}_2$ gauge and double-semion models with the \emph{double line} tensor network of Ref.~\cite{GuTopo}, where the authors also hint at an unpublished result for generic string-net models.

\section{\label{sec:stringnets}String-nets and the fat lattice}

String-net models were introduced in \cite{LevinWen} in order to encode
the universal physical properties of doubled $(2+1)D$ topological phases of
matter in quantum lattice models with few-body interactions. The simple
structure of their Hamiltonians reflects their conception as infrared fixed points
of renormalization group flows. Explicitly, these Hamiltonians are exactly solvable
because they are given by the sum of mutually commuting terms. In the following,
we consider those models in \cite{LevinWen} which exhibit a well-defined
continuum limit. 

These models are defined on a hexagonal lattice $\Lambda$. Local degrees of
freedom are associated with oriented edges of $\Lambda$ and elements
of the computational basis are labelled by $i\in\{1,\dots,N\}$. These
labels may be interpreted as particle species propagating along the edges.
For each label $i$ there is a unique label $i^*$ denoting its antiparticle,
and reversing the orientation of an edge corresponds to the mapping $i\mapsto i^*$.
The label $1$ stands for the absence of any particle (vacuum). Furthermore, each
instance of a string-net model is equipped with a set of {\em fusion rules}
$\delta_{ijk}$ specifying allowed ($\delta_{ijk}=1$) and forbidden
($\delta_{ijk}=0$) configurations of labels incident to a vertex. Given the set
of labels and their fusion rules, one can build a {\em tensor category}
which includes recoupling relations encapsulated in the symbol $F_{kln}^{ijm}$ (akin to
the $6j$-symbol in the theory of angular momentum), and an assignment of {\em quantum dimensions} $d_i$ to the labels. The {\em total quantum dimension} is given by $\mathcal{D}=\bigl(\sum_{i=1}^N d_i^2\bigr)^{1/2}$.

Fixed-point wavefunctions are constructed from local constraints \cite{LevinWen} which are
crafted so as to enforce topological invariance of the wavefunction. These
local constraints are assembled from the objects $d_i$, $F_{kln}^{ijm}$ introduced above.
Furthermore there is a natural correspondence between physical configurations on $\Lambda$
and configurations of string-nets in the {\em fat lattice}. The latter is constructed from
the physical lattice $\Lambda$ by puncturing the underlying surface at the center of each
plaquette. String-nets consist of oriented strings carrying labels in the set $\{1,\dots,N\}$,
joined at trivalent branching points in a way that respects the fusion rules $\delta_{ijk}$,
and avoiding the punctures. String-net configurations are defined to be equivalent if
they can be transformed into each other using the local relations (smooth deformations
avoiding punctures, recoupling by $F$-symbols, trading isolated loops for quantum
dimensions, and label conservation). Equivalence classes are identified with physical configurations. Note that the physical configuration itself can be regarded
as a particular string-net identical with the physical lattice. We will refer to this
particular string-net as the canonical representative of the equivalence class, and its uniqueness
is ensured by Mac Lane's coherence theorem \cite{MacLane}.

The Hamiltonian on the physical lattice reads
\begin{displaymath}
    H = -\sum_v A_v -\sum_p B_p,
\end{displaymath}
where the sums range over the vertices and plaquettes of the lattice. Vertex
terms are projectors enforcing the fusion rules
\begin{displaymath}
    A_v=\sum_{i,j,k\in v} \delta_{ijk}\,\ket{ijk}\bra{ijk}
\end{displaymath}
while plaquette projectors represent the kinetic part of the Hamiltonian and
are defined by
\begin{displaymath}
    B_p = \sum_{\alpha_p=1}^N \frac{d_{\alpha_p}}{\mathcal{D}^2} B_p^{\alpha_p},
\end{displaymath}
where $B_p^{\alpha_p}$ acts on the plaquette $p$ together with the outer legs of $p$.
Its precise definition is given in \cite{LevinWen}, as well as the
following simple graphical interpretation on the fat lattice: $B_p^{\alpha_p}$
creates an isolated loop of label $\alpha_p$ around the puncture at plaquette $p$.

\section{\label{sec:networks}Ground states of string-net models as tensor networks}

\begin{figure}
    \includegraphics{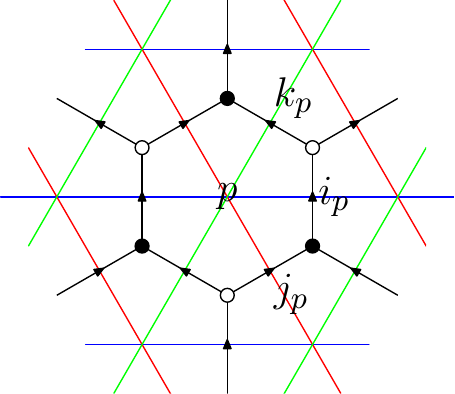}
    \caption{\label{fig:lattices} Hexagonal lattice $\Lambda$ and its dual
             $\Lambda^*$. Horizontal edges (blue) of $\Lambda^*$ belong to
             $E_1$, diagonal edges (red) to $E_2$ and diagonal edges (green) to
             $E_3$. The directed edges of $\Lambda$ are labelled by uniquely
             associating them to a face. Circled vertices of $\Lambda$ belong to
             the even sublattice $\Lambda_1$, filled ones to the odd sublattice
             $\Lambda_2$.}
\end{figure}

Let $\Lambda^*$ denote the dual lattice of $\Lambda$. It is
instructive to decompose the edge set of $\Lambda^*$ as
\begin{equation}
    E(\Lambda^*)=\bigcup_{i=1}^3 E_i
\end{equation}
where $E_1$ denotes the set of horizontal edges, $E_2$ one set of parallel diagonal edges,
and $E_3$ the other one as can be seen from Fig.~\ref{fig:lattices}.

\begin{figure*}
    \includegraphics{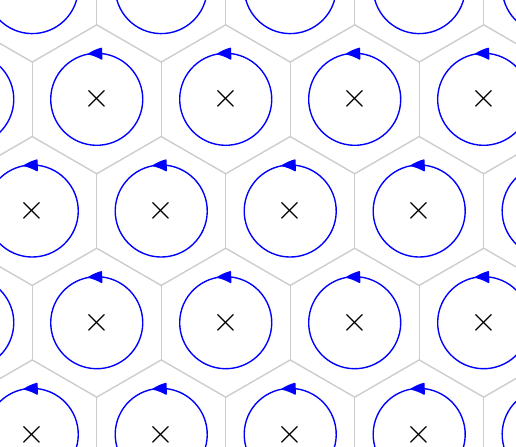}\hfill$\mapsto$\hfill
    \includegraphics{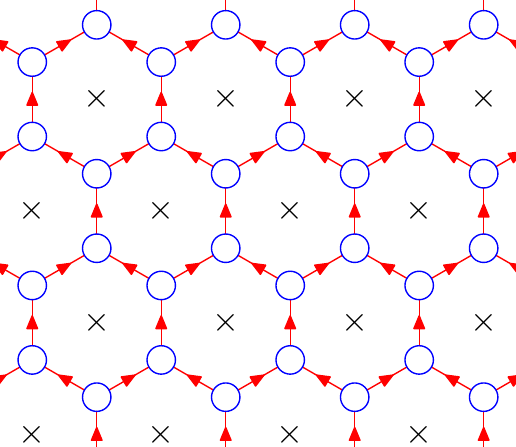}\hfill$\mapsto$\hfill
    \includegraphics{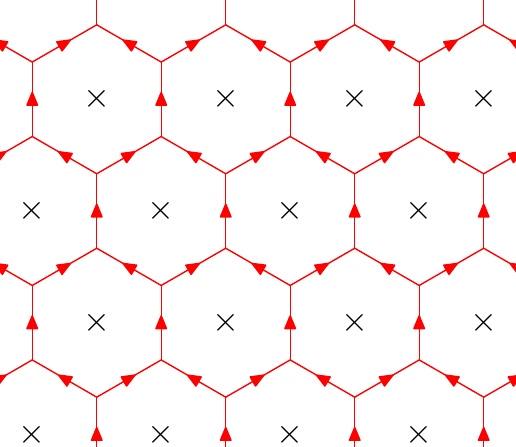}
    \caption{\label{fig:latticeReduction} Reducing the fat lattice. In (a)
             an isolated loop at face $p$ carries a label $\alpha_p$ and all
             edges $E(\Lambda)$ (grey) of the physical lattice are labelled by
             $0$. The overall quantum state is denoted by $\ket{\{\alpha_p\}}$.
             Three rounds of $F$-moves reduce to the fat lattice
             configuration (b) which carries both the final physical labels
             $i_p$, $j_p$, $k_p$ as well as the labels $\alpha_p$ at the vertex
             loops. This is denoted by $\ket{\{\alpha_p,i_p,j_p,k_p\}}$. In
             (c) each vertex has been fully reduced to the physical lattice by a
             sequence of another two $F$-moves each. The corresponding physical
             quantum state is given by $\ket{\{i_p,j_p,k_p\}}$.}
\end{figure*}

In order to construct an explicit graphical expression for a
ground state of a string-net model, we start with the state $\ket{1\dots 1}$
on the physical lattice where all edges carry the vacuum label $1$. Note
that this can be represented by a completely empty fat lattice. Obviously,
this state is an eigenstate of all the $A_v$ operators with eigenvalue $+1$.
Since the Hamiltonian is frustration-free we end up in the ground level
by applying the projection $\prod_p B_p$.

Thus, up to an overall factor, this ground state on the physical lattice
is represented by the following string-net state on the fat lattice:
\begin{equation}
	\label{eq:fatbaby}
    \ket{\Psi_0}=\biggl(\prod_p d_{\alpha_p}\biggr)\,\ket{\{\alpha_p\}}
\end{equation}
where $\ket{\{\alpha_p\}}$ denotes the string-net configuration shown
in Fig.~\ref{fig:latticeReduction}(a).

From now on we will use the local relations of the string-net model
in order to reduce Eq.~(\ref{eq:fatbaby}) to its canonical representative,
which can be directly translated into a configuration on the physical lattice.

After applying three rounds of recouplings involving $F$-symbols ($F$-moves)
to the strings on the fat lattice one has:
\begin{eqnarray}
    \ket{\Psi_0} &=& \sum_{\{\alpha_p\}}\biggl(\prod_p d_{\alpha_p}\biggr)
    \!\sum_{\{i_p,j_p,k_p\}}
    \!\biggl(\,\prod_{(p,q)\in E_1}
    \!F_{\alpha_q^*\alpha_qi_p}^{\alpha_p^*\alpha_p0}\biggr)\nonumber \\
    & &{}\times
    \biggl(\,\prod_{(p,q)\in E_2}
    \!F_{\alpha_q^*\alpha_qj_p}^{\alpha_p^*\alpha_p0}\biggr)
    \biggl(\,\prod_{(p,q)\in E_3}
    \!F_{\alpha_q^*\alpha_qk_p}^{\alpha_p^*\alpha_p0}\biggr)\nonumber \\
    & &{}\times\,\ket{\{\alpha_p,i_p,j_p,k_p\}},
\end{eqnarray}
where $\ket{\{\alpha_p,i_p,j_p,k_p\}}$ denotes the state of the fat
lattice as shown in Fig.~\ref{fig:latticeReduction}(b).
Using the normalization
\begin{equation}
    F_{j^*jk}^{ii^*0}=\sqrt{\frac{d_k}{d_id_j}}\,\delta_{ijk}
\end{equation}
this expression can be simplified in the case of an infinite or periodic lattice
to yield:
\begin{equation}
    \ket{\Psi_0}=\!\sum_{\{\alpha_p,i_p,j_p,k_p\}_\star}\!
    \biggl(\prod_p \frac{\sqrt{d_{i_p}d_{j_p}d_{k_p}}}{d_{\alpha_p}^2}\biggr)\,
    \ket{\{\alpha_p,i_p,j_p,k_p\}}.
\end{equation}
Note that we have
omitted the $\delta$-symbols and rather restricted the sum to configurations
$\{\alpha_p,i_p,j_p,k_p\}_\star$ that respect the branching rules of the
particular string-net model.

\begin{figure}
    \includegraphics{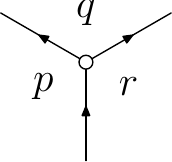}
    \includegraphics{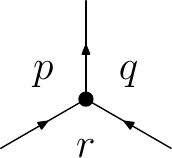}
    \caption{\label{fig:vertexLoops} Even (circled) and odd (filled) vertices of $\Lambda$ with
             their surrounding faces.}
\end{figure}

For a full reduction to the physical lattice we eventually need to remove the loops at the
vertices. This can be done by applying two $F$-moves at each vertex:
\begin{eqnarray}
    \ket{\{\alpha_p,i_p,j_p,k_p\}} &=&
    \biggl(\prod_p\frac{d_{\alpha_p}^2}{\sqrt{d_{i_p}d_{k_p}}}\biggr)
    \prod_{v\in \Lambda_1}f(v)\nonumber \\
    & &{}\times\!\prod_{v'\in \Lambda_2}\!g(v')\,\ket{\{i_p,j_p,k_p\}}
\end{eqnarray}
Here $\Lambda_i$ denote the even and odd sublattices of $\Lambda$ respectively
and furthermore one has:
\begin{eqnarray}
    f(v)&=&F_{k_r\alpha_qi_p}^{\alpha_p^*j_p\alpha_r}\\
    g(v')&=&F_{j_q^*\alpha_qk_p}^{\alpha_p^*i_p\alpha_r}
\end{eqnarray}
where the faces $\{p,q,r\}$ of $\Lambda$ surround an even vertex $v$ or an odd
vertex $v'$ as indicated in Fig.~\ref{fig:vertexLoops}.

At this point the ground state of the string-net model can be written in terms
of the physical lattice only:
\begin{eqnarray}
    \ket{\Psi_0} &= &\!\sum_{\{\alpha_p,i_p,j_p,k_p\}}\!
    \biggl(\prod_p\sqrt{d_{j_p}}\biggr)\prod_{v\in \Lambda_1}f(v)
    \prod_{v'\in \Lambda_2}g(v')\nonumber \\
    & &{}\times\,\ket{\{i_p,j_p,k_p\}}.
\end{eqnarray}
Note that because of the convention for the $F$-symbols in \cite{LevinWen} the branching rules at
each vertex are automatically satisfied and we no longer need to restrict the
sum. This allows one to isolate the basis coefficients:
\begin{equation}
    \lambda_{\{i_p,j_p,k_p\}}=\biggl(\prod_p\sqrt{d_{j_p}}\biggr)
    \sum_{\{\alpha_p\}}\prod_{v\in \Lambda_1}f(v)\prod_{v'\in \Lambda_2}g(v').
    \label{eq:coefficients}
\end{equation}
It is this very expression that we are now going to write in a graphical fashion
as a contracted tensor network.

\begin{figure*}
    \includegraphics{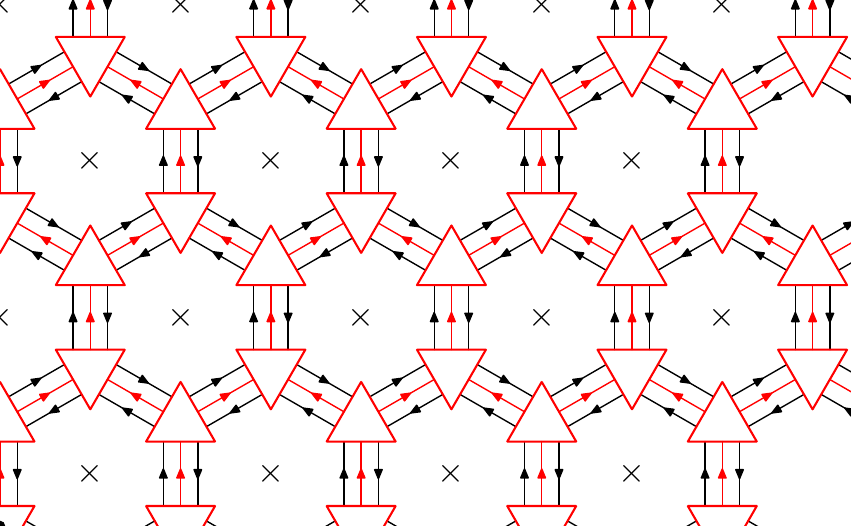}\hfill
    \includegraphics{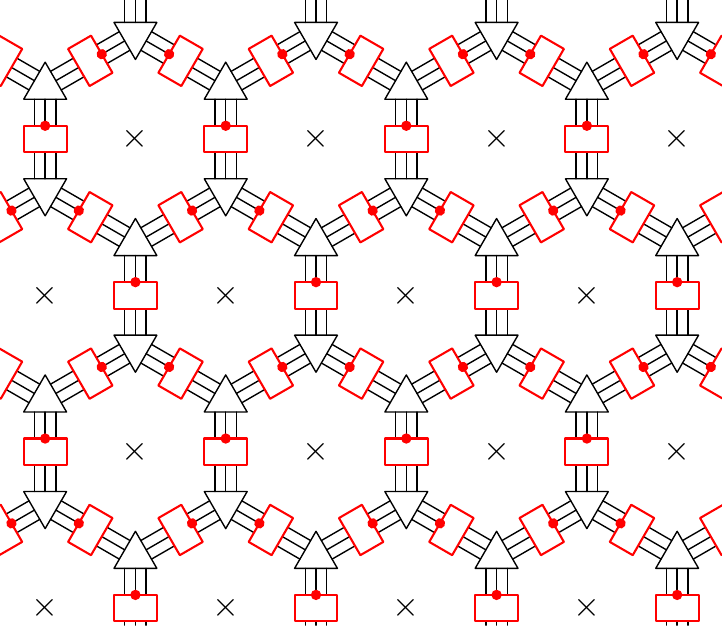}
    \caption{\label{fig:tensorNetwork1} (a) Tensor network describing the ground
             state of an arbitrary string-net model. In this figure only virtual
             bonds (black) are summed over while physical indices (red) are left
             uncontracted. (b) Tensor network describing the ground state of an
             arbitrary string-net model. In this figure all bonds are
             contracted.}
\end{figure*}

In order to write the coefficients of the string-net model ground state given by
Eq.~(\ref{eq:coefficients}) in a graphical fashion it is instructive to proceed
locally. Let us therefore consider an arbitrary face $a$ of $\Lambda$ together
with its next neighbours $b,\dots,g$. Obviously, the sum over $\alpha_a$ can now
be carried out immediately and will involve no more than six $F$-symbols. Thus
we obtain the following local expression:
\begin{eqnarray}
    \lambda_{\{i_p,j_p,k_p\}} &\sim&
    \sqrt{d_{j_a}d_{j_d}}\,\sum_{\alpha_a}
    F_{k_g\alpha_bi_a}^{\alpha_a^*j_a\alpha_g}
    \,F_{k_a\alpha_ci_d}^{\alpha_d^*j_d\alpha_a}
    \,F_{k_f\alpha_ai_e}^{\alpha_e^*j_e\alpha_f}\nonumber \\
    &&{}\times F_{j_c^*\alpha_ck_a}^{\alpha_a^*i_a\alpha_b}
    \,F_{j_d^*\alpha_dk_e}^{\alpha_e^*i_e\alpha_a}
    \,F_{j_a^*\alpha_ak_f}^{\alpha_f^*i_f\alpha_g}.
    \label{eq:localCoefficient}
\end{eqnarray}
Now define two sets of vertex tensors for the even and odd sublattices of
$\Lambda$ by
\begin{eqnarray}
    \vcenter{\hbox{\includegraphics{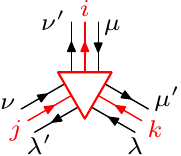}}}
    &:=& T_{\mu\mu'\nu\nu'\lambda\lambda'}^{[ijk]} \\
    \vcenter{\hbox{\includegraphics{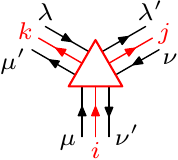}}}
    &:=& \tilde{T}_{\mu\mu'\nu\nu'\lambda\lambda'}^{[ijk]} 
\end{eqnarray}
where
\begin{eqnarray}
    T_{\mu\mu'\nu\nu'\lambda\lambda'}^{[ijk]}
    &:=& \sqrt{d_j}\,F_{k\mu i}^{\nu^*\!j\lambda}
    \,\delta_{\mu\mu'}\delta_{\nu\nu'}\delta_{\lambda\lambda'} \\
    \tilde{T}_{\mu\mu'\nu\nu'\lambda\lambda'}^{[ijk]}
    &:=& F_{j^*\lambda k}^{\mu^*i\nu}
    \,\delta_{\mu\mu'}\delta_{\nu\nu'}\delta_{\lambda\lambda'}
\end{eqnarray}
and contract them according to the network given in
Fig.~\ref{fig:tensorNetwork1}(a). If we cut out a single face of this network it
can easily be verified that it exactly reproduces the local form of our
coefficients as in Eq.~(\ref{eq:localCoefficient}), up to the factor
$\sqrt{d_{j_e}}$ (which can be absorbed, as the summation is extended to the
adjacent faces).

Thus we have obtained a simple graphical notation that describes the ground
state of an arbitrary string-net model and involves local terms only. In fact,
following the arguments of \cite{LevinWen}, our graphical calculus encompasses
the ground states of all ``doubled'' topological phases in the infrared limit.

We can also pull out the indices from the vertex tensors and collect physical
indices that denote particle and antiparticle into a single physical index at
the edge. This can be done by defining the following tensors:
\begin{equation}
    \vcenter{\hbox{\includegraphics{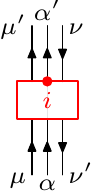}}}\!=
    A_{\alpha\alpha'\mu\mu'\nu\nu'}^{[i]}
\end{equation}

\begin{eqnarray}
    \vcenter{\hbox{\includegraphics{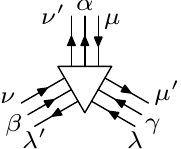}}}
    &=& B_{\alpha\beta\gamma\mu\mu'\nu\nu'\lambda\lambda'}\\
    \vcenter{\hbox{\includegraphics{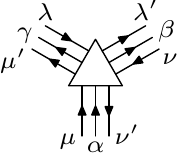}}}
    &=& \tilde{B}_{\alpha\beta\gamma\mu\mu'\nu\nu'\lambda\lambda'}
\end{eqnarray}
where
\begin{eqnarray}
    A_{\alpha\alpha'\mu\mu'\nu\nu'}^{[i]}
    &:=& \delta_{i\alpha}\delta_{\alpha\alpha'}\delta_{\mu\mu'}
    \delta_{\nu\nu'}\\
    B_{\alpha\beta\gamma\mu\mu'\nu\nu'\lambda\lambda'}
    &:=& T_{\mu\mu'\nu\nu'\lambda\lambda'}^{[\alpha\beta\gamma]}\\
    \tilde{B}_{\alpha\beta\gamma\mu\mu'\nu\nu'\lambda\lambda'}
    &:=& \tilde{T}_{\mu\mu'\nu\nu'\lambda\lambda'}^{[\alpha\beta\gamma]}
\end{eqnarray}
and contracting them according to Fig.~\ref{fig:tensorNetwork1}(b). Note that
the vertex tensors $T$ and $B$ only differ in how their indices are regarded:
what used to be a physical index of $T$ has been changed into a virtual one of
$B$. Thus the vertex tensors $B$ and $\tilde{B}$ are contracted on the virtual
level exclusively.

\vspace{1em}

\section{\label{sec:conclusion}Conclusions and outlook}

In this paper we have derived a remarkably simple tensor network representation for Levin
and Wen's string-net ground states. This construction follows directly from the
characterisation of these states as simultaneous $+1$ eigenstates of the
projectors in the Hamiltonian. It also heavily relies on the notion of the fat
lattice. Understanding string-net models in terms of the mapping from the fat lattice
to the physical lattice thus leads to insight and useful results. The tensor network is
built from the fusion rules
and $F$-tensors of the tensor category underlying the string-net model.

Note that from our Boltzmann weight tensor network one can trivially build a PEPS
representation. In the case of quantum double
models, which can be explicitly written as string-net models, dramatic
simplifications to this PEPS representation are possible due to
their group-theoretical properties. Also, for
a general string-net model it is possible to express excited states by absorbing
their corresponding open string operators into a ground state tensor network representation.
These topics will be discussed in \cite{QDSN}.

Due to its simplicity (as compared to the full-fledged specification of all stabilizer operators) this tensor network representation will help study a range
of properties of the ground level sector. For example, the topological entanglement entropy \cite{TEE} is one such property hinting at the presence of a new kind of multipartite, long-range entanglement that appears to underlie topological order. Along the same lines it may be interesting to establish criteria for a tensor network to represent a ground state of some topologically ordered quantum system. Of course, for this it would be necessary to extend the present analysis beyond the infrared limit as given by the string-net models.

\begin{acknowledgments}
We thank J.\,I.~Cirac and N.~Schuch for valuable and inspiring discussions.
G.~Vidal acknowledges support from Australian Research Council (FF0668731, DP0878830).
\end{acknowledgments}

\end{document}